\input{aipcheck}

\documentclass[
    ,final            
  ]
  {aipproc}
\usepackage{epsfig}
\layoutstyle{6x9}

\begin{document}
\title{Spin Sum Rules at Low $Q^2$}
\author{Jian-ping Chen}{
  address={Thomas Jefferson National Accelerator Facility, Newport News, 
VA 23606, USA}}

\begin{abstract}
Recent precision spin-structure data from Jefferson Lab have significantly 
advanced our knowledge of nucleon structure at low $Q^2$. 
Results on the neutron spin sum rules and polarizabilities in the 
low to intermediate $Q^2$ region are presented. The Burkhardt-Cuttingham 
Sum Rule was verified within experimental uncertainties.
When comparing with theoretical calculations, results on spin
polarizability show
surprising disagreements with Chiral Perturbation Theory predictions. 
Preliminary results on first moments at very low $Q^2$ are also presented.
\end{abstract}
\maketitle
\subsection{Introduction}
\label{intro}

Sum rules involving the spin structure of the nucleon offer an important 
opportunity to study QCD. In recent years
the Bjorken sum rule~\cite{Bjorken} at large $Q^2$ and 
the Gerasimov, Drell and Hearn (GDH) sum rule~\cite{gdh} at $Q^2=0$
have attracted large experimental and theoretical~\cite{spin} efforts that have provided us with rich information. 
A generalized GDH sum rule~\cite{ggdh} connects the GDH sum rule with the 
Bjorken sum 
rule and provides a clean way to test theories with experimental data over the
entire $Q^2$ range. 
Spin sum rules relate the moments of the spin-structure functions to the 
nucleon's static properties or
real or virtual Compton amplitudes, which can be calculated theoretically. 
Refs.~\cite{chen05,dre} provide comprehensive reviews on this subject.

\subsection{Results on moments of the neutron spin-structure functions}
\par
Recently, the high polarized-luminosity available at Jefferson 
Lab has allowed a
study of nucleon spin structure with 
an unprecedented precision.  
The neutron results on both $g_1$ and $g_2$ from
Hall A were extracted from data on a $^3$He target polarized 
in both longitudinal and transverse directions. 

\begin{figure}[!hbt!]
\begin{minipage}[t]{8 cm}
\epsfig{file=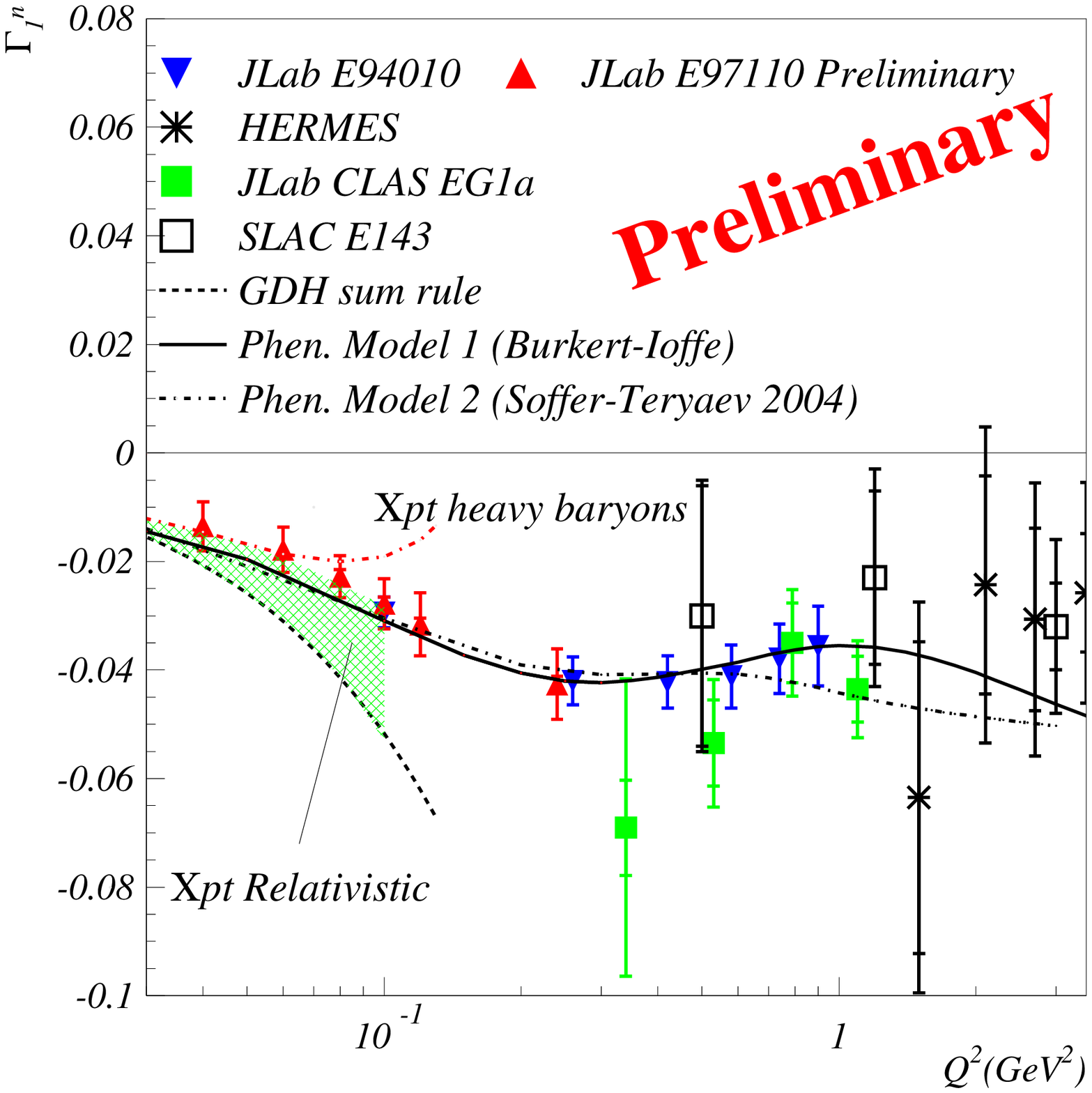,scale=0.35}
\end{minipage}
\begin{minipage}[t]{8 cm}
\vspace{-2.3in}
\epsfig{file=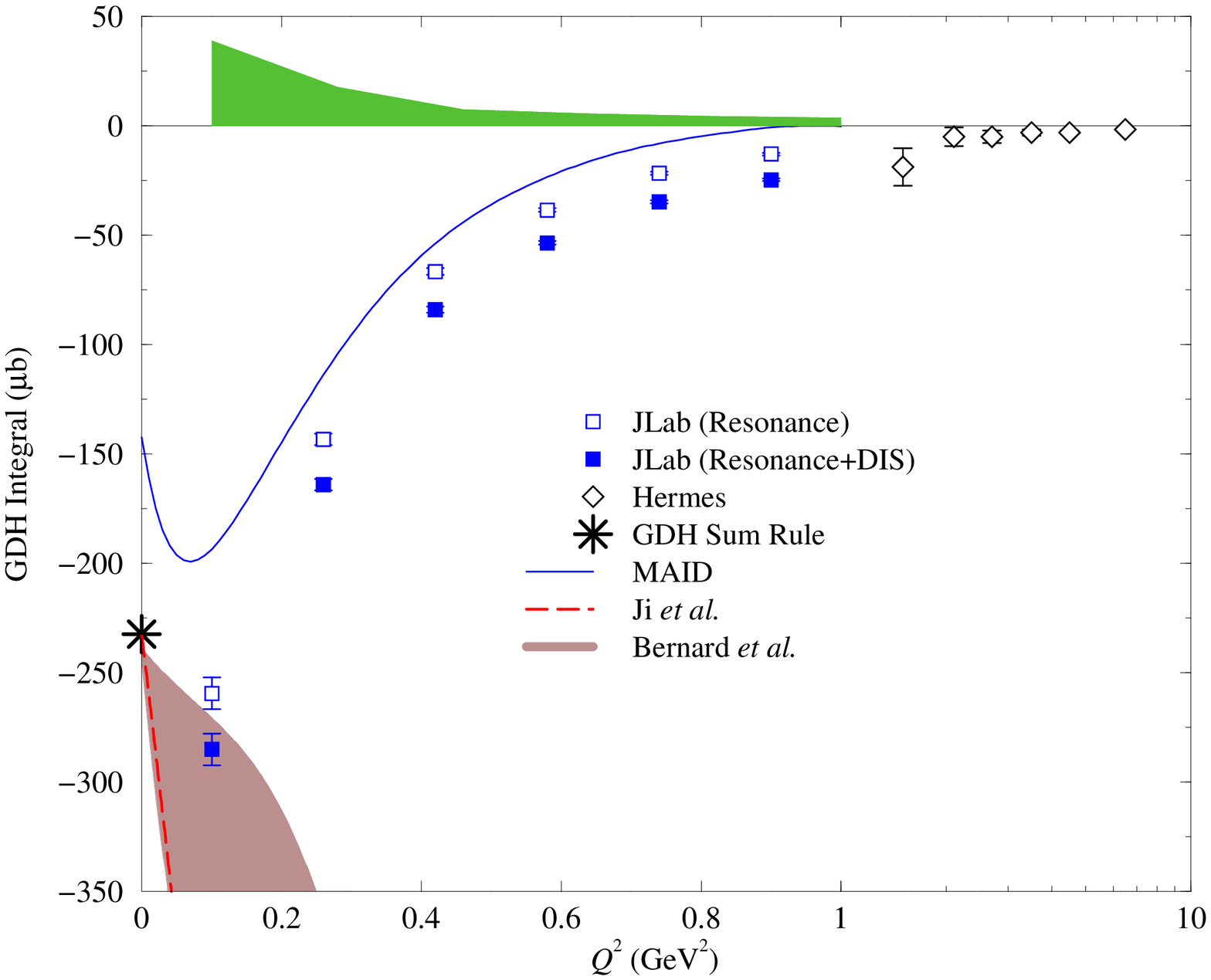,scale=0.35,angle=-90}
\vspace{-0.1in}
\end{minipage}
\begin{minipage}[t]{16.5 cm}
\caption {Results of  $\Gamma_1$ and GDH sum $I(Q^2)$ for the neutron~\protect\cite{e94010}.
The results are compared with $\chi$PT calculations of 
ref.~\protect\cite{chpt} (dashed line) and ref.~\protect\cite{chpt2} 
(shaded band).  
The MAID model calculation of ref.~\protect\cite{dre01},
is represented by a solid line.  Data from HERMES~\protect\cite{HERMES} are also shown.}
\label{fig:GDH}
\end{minipage}
\end{figure}

Fig.~\ref{fig:GDH} shows $\Gamma_1$~\cite{e97110,e94010} (left), 
the first moment of 
$g_1$, and the extended GDH integrals~\cite{e94010} (right) 
$I(Q^2)=\int_{\nu_{th}}^\infty [\sigma_{1/2}(Q^2)-\sigma_{3/2}(Q^2)]d\nu/\nu$ 
for the neutron. 
The left panel shows the preliminary results of $\Gamma_1^n$ at very low 
$Q^2$~\cite{e97110} together with the results at low to intermediate 
$Q^2$ region~\cite{e94010}. Also shown are the neutron results extracted from 
the deuteron and proton data from Hall B~\cite{eg1a} and high $Q^2$ data 
from HERMES~\cite{HERMES} and SLAC~\cite{SLAC}.
At $Q^2$=0, the GDH sum rule predicts the slope of $\Gamma_1$ (dotted lines). 
The behavior at low $Q^2$ can be calculated with Chiral Perturbation Theory
($\chi$PT). We show 
a Heavy Baryon $\chi$PT (HB$\chi$PT) calculation~\cite{chpt} 
(dashed lines) and a Relativistic Baryon $\chi$PT 
(RB$\chi$PT) calculation\cite{chpt2}
including vector mesons and $\Delta$ contributions (shaded band).
The predictions are in reasonable agreements with the data at the lowest $Q^2$
settings of 0.04 - 0.1 GeV$^2$.
At moderate to large $Q^2$ data are compared with two model 
calculations~\cite{sof02,bur92}. Both models agree well with the data.  

The open symbols on the right plot are measured GDH integral from 
pion threshold to $W=2$ GeV.   
The solid squares include an estimate of the unmeasured high-energy part.
The results indicate a smooth variation of $I(Q^2)$  to increasingly negative 
values as $Q^2$ varies from 0.9 GeV$^2$ towards zero.
The data (open squares) are more negative than the MAID model 
calculation\cite{dre01}. The GDH sum rule 
prediction, $I(0)=-232.8\,\mu{\rm b}$, is indicated in Fig.~\ref{fig:GDH}, 
along with  
extensions to $Q^2>0$ using the next-to-leading order HB$\chi$PT) 
calculation~\cite{chpt} (dashed line) and the RB$\chi$PT) 
calcualtion~\cite{chpt2} (shaded band) including resonance 
effects~\cite{chpt2}.

Combining the neutron results with the proton data from Hall B,
results on the moment of $g_1^p-g_1^n$, 
the generalized Bjorken sums~\cite{bjsum}, were obtained.
The data at high $Q^2$ values were used
to test the Bjorken sum rule as one of the fundamental tests of QCD.
They were also used to extract a value of strong coupling constant, $\alpha_s$.
The new JLab data at low $Q^2$ provide important information in the low 
energy region, where the strong interaction is non-perturbative.
An attempt~\cite{alphas} was made to extract an effective strong coupling, 
$\alpha_s^{eff}$ in the low $Q^2$ region. The extracted $\alpha_s^{eff}$ 
shows a trend of weakening $Q^2$-dependence with decreasing $Q^2$.  

\begin{figure}[!hbt!]
\begin{minipage}[t]{8 cm}
\vspace{-3.0in}
\epsfig{file=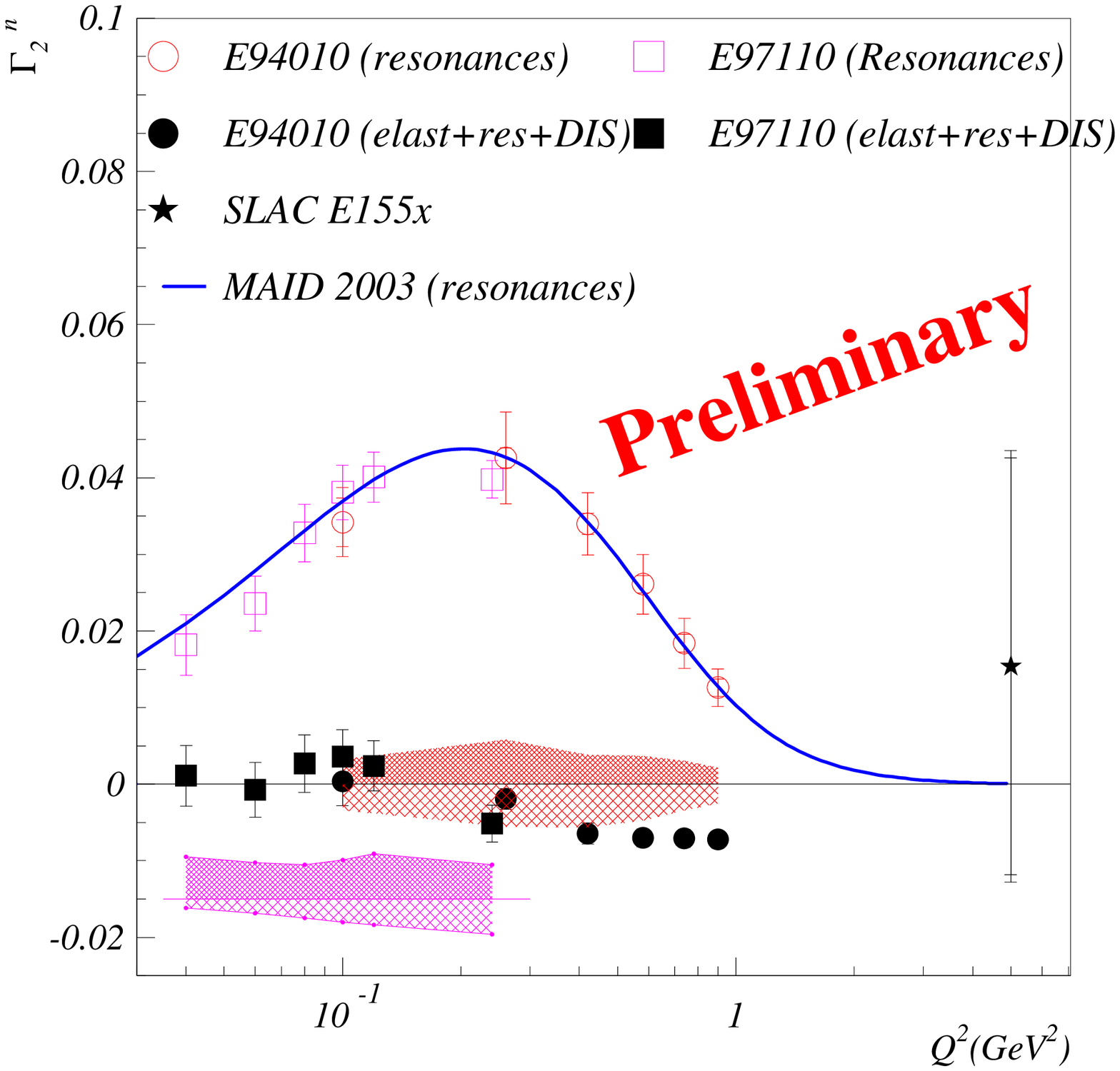,scale=0.35}
\end{minipage}
\begin{minipage}[t]{8 cm}
\epsfig{file=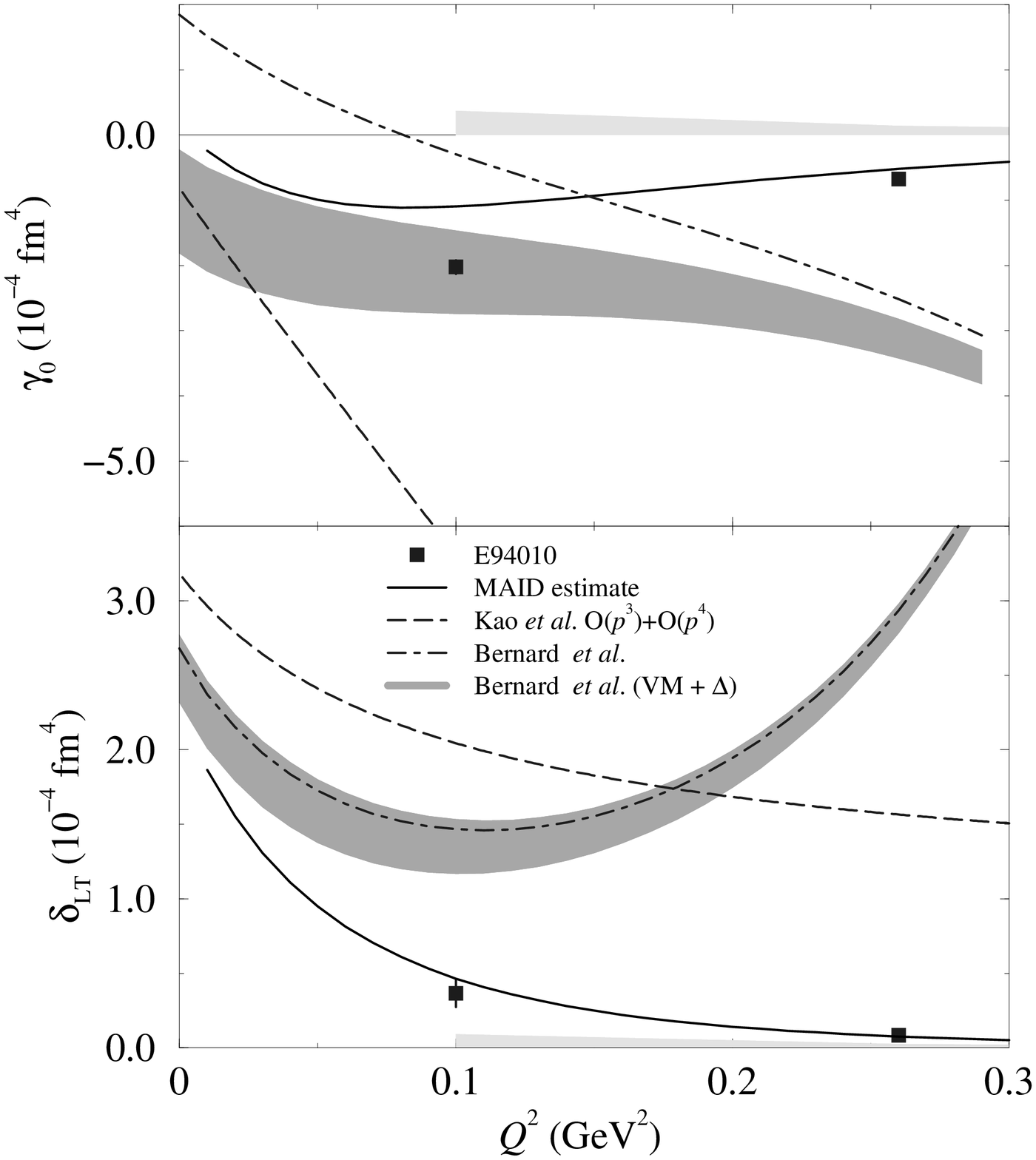,scale=0.35}
\end{minipage}
\begin{minipage}[t]{16.5 cm}
\vspace{-0.2in}
\caption {Results on the BC sum, $\Gamma_2^n (Q^2)$ (left)
at very low $Q^2$~\protect\cite{e97110} (preliminary) and at low to 
intermediate $Q^2$~\protect\cite{e94010}. Results on
the spin polarizabilities $\gamma_0^n$ (top-panel) and $\delta_{LT}^n$
(bottom-panel) at low $Q^2$~\protect\cite{e94010}. 
Solid curves are the MAID model~\protect\cite{dre01}
calculations. The dashed curves represent the heavy baryon$\chi $PT 
calculation~\protect\cite{chpt1}. 
The dot-dashed curves and the shaded bands represent the relativistic baryon
$\chi $PT calculation 
without and with~\protect\cite{chpt2} the $\Delta $ and vector meson 
contributions, respectively.
 }
\label{fig:BCsum}
\end{minipage}
\end{figure}

Preliminary results of the first moment of $g_2^n$, $\Gamma_2^{n}$, at very 
low $Q^2$ are plotted on the left panel of 
Fig.~\ref{fig:BCsum} in the measured region (open squares). Solid suqares 
show the results after adding an elastic and an estimated low-$x$
contributions. Also shown as open circles (measured) and solid circles
(total) are the previously published results at low $Q^2$ to intermediate 
$Q^2$. 
The MAID estimate agrees well with the measured resonance data. The two bands correspond to the experimental systematic
errors and the estimate of the systematic error 
for the low-$x$ extrapolation. The total results are consistent with the BC 
sum rule~\cite{BC}. 
The SLAC E155x collaboration\cite{SLAC} previously reported 
a neutron result at high $Q^2$ (star) with a rather 
large error bar. On the other hand, the SLAC proton result was reported to 
deviate from the BC sum rule by 3 standard deviations. 

The generalized spin polarizabilities provide benchmark tests
of $\chi$PT calculations at low $Q^2$.
Since the generalized polarizabilities have an extra $1/\nu^2$ 
weighting compared to the first moments, these integrals have 
less contributions from the large-$\nu$ 
region and converge much faster, which minimizes the uncertainty due to
the unmeasured region at large $\nu$. 
At low $Q^2$, the 
generalized polarizabilities have been evaluated with next-to-leading order 
$\chi$PT 
calculations~\cite{chpt1,chpt2}.
Measurements of the generalized spin
polarizabilities are an important step in understanding the dynamics of
QCD in the chiral perturbation region.

The results for $\gamma_0(Q^2)$~\cite{e94010} are shown
in the top-right panel of Fig.~\ref{fig:BCsum}. 
The data are compared with 
a next-to-leading order ($O(p^4)$) HB$\chi$PT 
calculation~\cite{chpt1}, a next-to-leading order RB$\chi$PT
calculation and the same calculation explicitly including 
both the $\Delta$ resonance and vector meson contributions~\cite{chpt2}.
Predictions from the MAID model~\cite{dre01} are also shown.
At the lowest $Q^2$ point,
the RB$\chi$PT
calculation including the resonance contributions
is in good agreement with the experimental result.
For the HB$\chi$PT calculation without explicit resonance contributions, 
discrepancies are large even at $Q^2 = 0.1$ GeV$^2$. 
This might indicate the significance of the resonance contributions.
The data are in reasonable agreement with the MAID predictions.
Since $\delta_{LT}$ is 
insensitive to the
$\Delta$ resonance contribution, it was believed that $\delta_{LT}$ should be
more suitable than $\gamma_0$ to serve as a testing ground for the chiral 
dynamics of QCD~\cite{chpt1,chpt2}.
The bottom-right panel of Fig.~\ref{fig:BCsum} shows 
$\delta_{LT}$~\cite{e94010} compared to
$\chi$PT calculations and the MAID predictions. While the MAID predictions are in good agreement with the results, it is surprising to see
that 
the data are in significant disagreement with the $\chi$PT calculations 
even at the lowest $Q^2$, 0.1 GeV$^2$. 
This surprising disagreement (``$\delta_{LT}$ puzzle'') presents a 
significant challenge to the present Chiral Perturbation Theory.

The spin polarizabilities data at very low $Q^2$~\cite{e97110} 
should be avalaible soon. 
These results will provide benchmark tests to the
$\chi$PT calculations at the kinematics where they are expected to work.
A new proposal~\cite{e08027} was recently approved to measure $g_2^p$ with 
a transversely polarized proton target in the low $Q^2$ region. It will 
provide an isospin separation of the spin polarizabilities to shed light 
on the ``$\delta_{LT}$'' puzzle. 

\subsection{Summary}
In summary, the high polarized-luminosity available at
JLab has provided us with high-precision nucleon spin structure data
in the low to intermediate $Q^2$ region. These data help to 
study the non-perturbative region and the transition 
between perturbative and non-perturbative regions of QCD.
\medskip

\footnotesize{
The work presented was supported in part 
by the U. S. Department of Energy (DOE)
contract DE-AC05-84ER40150 Modification NO. M175,
under which the
Southeastern Universities Research Association operates the 
Thomas Jefferson National Accelerator Facility.
}

\end{document}